# Symbolic Dynamics and Markov Partitions for the Stadium Billiard

Kai T. Hansen*
Fakultät für Physik, Universität Freiburg
Hermann-Herder-Strasse 3, D-79104 Freiburg, Germany
e-mail: k.t.hansen@fys.uio.no
and
Predrag Cvitanović
Center for Chaos and Turbulence Studies, Niels Bohr Institute,
Blegdamsvej 17, DK-2100 Copenhagen Ø, Denmark

February 6, 1995

**Abstract**

We investigate the Bunimovich stadium dynamics and find that in the limit of infinitely long stadium the symbolic dynamics is a subshift of finite type. For a stadium of finite length the Markov partitions are infinite, but the inadmissible symbol sequences can be determined exactly by means of the appropriate pruning front. We outline a construction of a sequence of finite Markov graph approximations by means of approximate pruning fronts with finite numbers of steps.

## 1 Introduction

Good symbolic dynamics is a prerequisite to analysis of the dynamics of chaotic systems. For one-dimensional mappings the theory is well developed:

*Also at: Physics Department, University of Oslo, Box 1048, Blindern, N-0316 Oslo, Norway.

for example, the kneading theory of Milnor and Thurston[1] yields a complete description of admissible orbits of unimodal maps. In higher dimensions the Smale horseshoe[2] is an example of a class of dynamical systems described by a complete binary symbolic dynamics. Some progress has also been made in the description of generic smooth two-dimensional maps. For example, it has been conjectured that the admissible orbits of the Hénon map can be described by a subset of binary symbol sequences by means of the so called *pruning fronts*[3, 4, 5, 6]. Different billiard systems have also been investigated, and some of them can be exactly described by suitable pruning fronts[7, 8]. Symbolic dynamics is closely related to Markov partitions of the dynamical phase space; crudely speaking, an alphabet labels the distinct regions of the phase space, and a Markov graph or a transition matrix indicates how these are interconnected by the dynamics. Systems which can be described by simple symbolic dynamics, such as the horseshoe, the well separated 3-disk pinball[9], and the cat map[10], have simple Markov partitions. For a variety of more generic ergodic billiards it has been proved by Sinai, Bunimovich and others[11, 12, 13] that there exist infinite but countable Markov partitions. Such systems can be approximated by more and more refined finite Markov partitions, and correspondingly more and more complicated symbolic dynamics.

In this paper we give an example of such procedure by constructing a symbolic dynamics for the Bunimovich stadium[14]. The first such symbolic dynamics was introduced by Biham and Kvale[15]; ours is essentially continuation of their work, resulting in a more compact, desymmetrized covering symbolic dynamics. Meiss[16] has offered a rather different classification of a subset of stadium billiard orbits by their rotation numbers. While this is the natural labeling scheme in the integrable limit of collapsing the stadium into a circle, Biham-Kvale's and our symbolic dynamics is natural in the long stadium limit. Also, in contrast to the rotation number labeling, our symbolic dynamics labels all orbits. The main new results presented here are a complete Markov partition for the infinitely long stadium and construction of a pruning front that gives the exact description of all admissible orbits for a given finite length stadium.

The paper is organized as follows: In sect. 2 we describe the stable/unstable manifold structure for the stadium billiard. In sect. 3 we first describe a variant of the Biham-Kvale symbolic dynamics, and then introduce a more compact symmetry reduced covering symbolic dynamics for an 1/4 stadium of infinite length. In sect. 4 we explain the relation between this symbolic dynamics and Markov partitions of the phase space, and in sect. 5 we use the topology of such Markov partitions to construct a topologically faithful symbol plane representation of the dynamics. This enables us to construct the exact pruning front for a given finite length stadium in sect. 6. We then show how these pruning fronts can be approximated by finite grammar cycle expansions, and apply these to estimate the topological entropy of a finite length stadium.

## 2 Stable/unstable manifolds

The stadium billiard consists of a point particle moving freely within a two dimensional domain, reflected elastically at the border which consists of two semi-circles of radius 1 connected by two straight walls of length $2a$. At the points where the straight walls meet the semi-circle, the curvature of the border changes discontinuously; these are the only singular points on the border. The length $a$ is the only parameter, and we are interested in studying the dynamics at different values of $a$.

Typical structure of the stable/unstable manifolds structure of orbits is illustrated by the unstable period-2 cycle manifolds at $a = 1$ drawn in Fig. 1. We draw the manifolds in the Poincaré map $(x, \phi)$, where $x$ is a distance along the border to the bounce point and $\phi$ is the outgoing angle of the bounce. Alternatively, we could have used the $(x, \cos\phi)$ area preserving coordinates, but for the discussion here this is not important. The unstable manifolds for the straight wall phase space are drawn in Fig. 1 (a), and for the semi-circle phase space in Fig. 1 (b). In Figs. 1 (c) and (d) both the stable and the unstable manifolds are drawn.

All intersections of the stable and the unstable manifolds for this hyperbolic map are transverse (at non-zero angles). In this type of billiard the smooth sections of the manifolds end at points with discontinuous derivative. The manifolds are folded with a sharp corner, a *turning point*, when they hit a singular point, because the derivative of the reflection of point particle is discontinuous here. However, the limit rays from each side of the singularity converge to the same orbit, so the manifolds are continuous, but with sharp breaks.

The stable and the unstable manifolds intersections, Fig. 1 (c) and (d),

yield the orbits homoclinic to this period-2 cycle. These are dense in the phase space. As $a$, the length of the stadium decreases, the manifolds move in such a way that infinite families of the stable/unstable manifolds intersections are lost. This occurs at the turning points of the manifolds, and this point where a smooth section is touching a turning point is the analogue of a homoclinic tangency for a smooth dissipative diffeomorphism. Newhouse has shown that for a parameter sweep through a homoclinic tangency of a smooth diffeomorphism there will also occur an infinity of stable orbits[17]. For a stadium any finite change of the parameter $a$ sweeps through bifurcations of an infinite number of periodic orbits, but no stable orbits occur. A more appropriate dissipative map analogue in this case is the Lozi map[18] whose stable/unstable manifolds also have sharp turning points.

The turning points break up the manifolds into infinite chains of piecewise smooth folds. The length of a smooth section may be arbitrarily short and this leads to difficulties in proving the ergodicity of the system and constructing Markov partitions[14]. Our strategy is to single out one family of turning points which we call the *primary* turning points; this family is the 2-$d$ billiards analogue of the critical point of a unimodal mapping. All other turning points are then (pre-)images of the primary ones. We then map the primary turning points into a symbol plane (defined below), and refer to this set of points as the *pruning front*. The area on one side of the pruning front corresponds to symbol sequences that are inadmissible for the given finite length stadium. These sequences are "pruned"; what remains is the set of all admissible symbol sequences for the system. Prerequisite to implementing this procedure is existence of a *covering* symbolic dynamics, such that not more than one dynamical orbit correspond to a given symbol string, but there may be many symbol strings that correspond to no dynamical orbits.

# 3 Covering symbolic dynamics

The covering symbolic dynamics has to allow for the description of all orbits, including those which, for a given stadium length $a$, lie on the forbidden side of the turning points of the manifold. These are orbits which exist for a sufficiently long stadium, but not necessarily for small $a$. The covering symbolic dynamics may also include symbol strings which *never* correspond to a dynamical orbit. The first example of such symbolic dynamics for the stadium was given by Biham and Kvale[15]. They associate $t$-th bounce of an orbit with 6-letter alphabet $S_t \in \{0,1,2,3,4,5\}$ defined as follows (see Fig. 2):

$$S_t = \begin{cases} 0 & \text{a bounce off the bottom wall.} \\ 1 & \text{a clockwise bounce off the left semi-circle or} \\ & \text{a single anticlockwise bounce off the left semi-circle.} \\ 2 & \text{a bounce off the upper wall.} \\ 3 & \text{an anticlockwise bounce off the right semi-circle or} \\ & \text{a single clockwise bounce off the right semi-circle.} \\ 4 & \text{a not single anticlockwise bounce off the left semi-circle.} \\ 5 & \text{a not single clockwise bounce off the right semi-circle.} \end{cases} \tag{1}$$

A bounce off a semi-circle is a single bounce if both the preceding and the following bounces are not off the same semi-circle. Biham and Kvale had tested this definition numerically and found that for all orbits they have tried no two different periodic orbits were described by the same symbol string. They showed that the periodic orbits which only exist for large values of $a$ can also be found numerically for small $a$ if one allows for bounces in the semi-circle inside the stadium and bounces off the straight walls outside the stadium. They also gave the "geometric" pruning rules, *ie.* a description of symbol strings that correspond to orbits that never exist in stadium of any length.

Here we reduce the symmetry and reduce the number of symbols in two steps. First we choose a slightly different 5-letter alphabet $\sigma_t \in \{0,1,2,3,4\}$ (see Fig. 3):

$$\sigma_t = \begin{cases} 0 & \text{the bounce is the first bounce in a semi-circle} \\ & (\textit{ie.}, \text{the previous bounce was not in the same semi-circle}). \\ 1 & \text{the bounce is clockwise in a semi-circle,} \\ & \text{but not the first bounce in this semi-circle.} \\ 2 & \text{the bounce is anticlockwise in a semi-circle,} \\ & \text{but not the first bounce in this semi-circle.} \\ 3 & \text{the bounce is off the bottom wall going right} \\ & \text{or off the top wall going left.} \\ 4 & \text{the bounce is off the bottom wall going left} \\ & \text{or off the top wall going right.} \end{cases} \tag{2}$$

A finite Markov graph is constructed from a finite list of subsequences which can never occur[5]; in this case there are four such rules. An orbit cannot have two consecutive bounces off the same straight wall or change the sense of direction along stadium between two straight wall bounces. This forbids the substrings $_33_$ and $_44_$. From the definitions it follows that the symbol 1 can only follow after a symbol 1 or 0, and the symbol 2 can only follow after a symbol 2 or 0. A Markov graph that excludes these substrings is drawn in Fig. 4. This graph does allow symbol dynamics fixed points $\overline{1} = 1^\infty$ and $\overline{2} = 2^\infty$, which do not exist in a stadium, but are approached by the "whispering gallery" periodic orbits with subsequences of $_1^k_$, $_2^j_$ bounces of arbitrary length. Such families of orbits, as well as the orbits with $_3^k_$, $_4^j_$ subsequences which approach the "bouncing ball" orbits, have positive but arbitrarily small Lyapunov exponents and require special treatment in applications such as computation of semiclassical spectra. The fixed point $\overline{0}$ corresponds to the unstable 2-cycle along the stadium, so the $\sigma_t$ reduced symbolic dynamics corresponds to a 1/2 stadium.

### 3.1 Desymmetrized symbolic dynamics

The symmetry of the Markov graph in Fig. 4 is due to the $C_{2v}$ symmetry of the stadium itself. Following ref. [19] we now reduce the symbolic dynamics to that appropriate to the fundamental domain, *ie.* the 1/4 stadium. The main idea is to relabel the trajectories so instead of keeping track of the labels for individual boundary segments we label the types of transitions between bounces.

There is a symmetry between symbol 1 and 2 and between 3 and 4, but we cannot simply identify these pairs of symbols. For example, it is clear by inspection that the string $101\cdots$ corresponds to an orbit topologically distinct from that labeled by the string $102\cdots$. The first string corresponds to an orbit which keeps the clockwise rotation, while the second string corresponds to an orbit which changes the sense of rotation from clockwise to anticlockwise. However, the strings $101\cdots$ and $202\cdots$ yield orbits which are identical except for a reflection or time reversal. Hence in defining new symbols we have to control the sense of rotation. In table 1 the symmetry reduced alphabet $s_t$ is defined from the two symbol combinations $\sigma_{t-1}\sigma_t$. Since the symbol 0 does not distinguish between a clockwise or an anticlockwise bounce, we also have to keep track of the symbol preceding a string of repeated 0's in order to decide which new symbol to use.

In applications it might be more convenient to use infinite alphabets[20] by lumping repeats of symbol $a$'s and $f$'s together with the preceding symbol into a single symbol, $_sa^k_ \rightarrow _s_{(k-1)}_$ etc., but we shall not do that here.

The Markov graph for the symmetry reduced alphabet $s \in \{a, b, c, \cdots, k\}$ is drawn in Fig. 5. This graph implements the "geometric" pruning rules, *ie.* excludes symbol sequences that cannot occur for any value of $a$, and provides a *covering* symbolic dynamics for the stadium, the starting point for our analysis of the finite length billiards. In contrast to the Markov graph of Fig. 4, here the symbols label paths from one node to the next, and there can be several paths connecting the nodes. The 3 nodes correspond to bounces either off the straight wall (I), off the semi-circle as a first bounce in the semi-circle (II), or off the semi-circle as a second or later bounce (III). Though it might seem that going from a 5-letter alphabet to an 11-letter alphabet is only a complication, the contrary is true: desymmetrization factorizes and simplifies the associated zeta functions and Fredholm determinants, and greatly improves convergence of computations over chaotic sets[19].

# 4 Markov partitions

In order to develop better intuition about this symbolic dynamics and its applicability to stadia of finite length $a$, we first explain how this Markov graph relate to a Markov partition of the phase space.

To each of the three nodes in Fig. 5 corresponds a partition of the three different parts of the phase space. This partition is obtained by letting the arrows into and out from the node define an area in the phase space which we shall call a *rectangle*. The points in a rectangle $s_0.s_1s_2$ correspond to orbits containing a subsequence $\cdots s_0.s_1s_2\cdots$, where all explicitly indicated $s_t$ symbols are fixed.

Node I corresponds to bounces off the straight wall, with a position $-a < x < a$ (with $x = 0$ on the center of the straight wall), and an outgoing angle $0 < \phi < \pi/2$. There are 5 ways to enter the node and 2 ways to leave. This yields a partition into 10 rectangles in the phase space I, Fig. 6.

Node II corresponds to a first bounce off the semi-circle with a position $0 < x < \pi$ (with the two singular end points at $x = 0, x = \pi$) and an outgoing angle $x/2 < \phi < \pi/2 + x/2$. There are 3 ways to enter the node and 5 ways to exit. This yields a partition of the phase space into 15 rectangles. Because we do not know the orientation of the preceding bounces, the past orientation doubles the partition of the phase space II into 30 rectangles, Fig. 7.

Node III corresponds to a second or later bounce off the semi-circle with a position $0 < x < \pi$, and an outgoing angle $0 < \phi < x/2$. This is the part of the phase space for the semi-circle not covered by the node II. There are 3 ways to enter this node and 4 ways to exit, yielding a partition of the phase space III into 12 rectangles, Fig. 8.

In the infinite length limit $a \to \infty$ this partitioning of the stadium phase space is complete. As $a$ decreases the rectangles decrease (in the sense of containing fewer orbits, not in any metric sense), and for sufficiently small parameter values some rectangles may be completely lost. This loss of orbits will be described below in terms of the pruning front.

The borders between the rectangles drawn in Figs. 6, 7 and 8 are found numerically. Most of these partition curves consist of the points from where an orbit goes to or came from the singular point of the wall after the required intenerary. Other curves are the points with a specific intenerary belonging to the stable or unstable manifold of the hyperbolic period-2 cycle. We indicate former borders by solid curves, and the latter by dashed curves. These curves partition the phase space into rectangles. Each rectangle in the phase space is labeled by the two symbols $s_{t-1}.s_t$.

The phase space partition I, Fig. 6, is drawn by noting that the five arrows going into the node I in Fig. 5 are $j.$, $h.$, $i.$, $k.$ and $f.$, and that the two outgoing arrows are $.g$ and $.f$. The partition curve between $.g$ and $.f$ is given by the orbits bouncing straight into the singular point, and the partition curve between $f.$ and $k.$ is given by orbits arriving directly from the singular point. The partition curves between $k.$ and $i.$ and between $h.$ and $j.$ are the given by orbits arriving at a point on the semi-circle wall from the singular point and then bouncing once off this semi-circle before reaching the straight wall. The partition curve between $i.$ and $h.$ is a part of the unstable manifold of the hyperbolic period-2 cycle (dashed curve). Points on this partition curve correspond to the orbits which hit the straight wall segment after having bounced an infinite number of times from one semi-circle to the other.

In phase space partition II, Fig. 7, three arrows $g.$, $c.$, and $b.$ go into the node II in Fig. 5, and there are five outgoing arrows $.d$, $.h$, $.b$, $.i$, and $.e$. If $\phi < x/2$ or $\phi > \pi/2 + x/2$, then the previous bounce was off the same semi-circle, *ie.* this bounce is not the first bounce off the semi-circle, and thus not belonging to phase space II. The partition curves between the rectangles are given by orbits going into the singular point directly, or after one bounce in the other semi-circle. The partition curves between the symbols $.d$ and $.h$ and between $.i$ and $.e$ are simply the straight lines $\phi = \pi/2 - x/2$, $\phi = \pi - x/2$. In addition to these partition curves we have drawn the first smooth fold of the unstable manifold of the 2-cycle which has a point in the center of this plane at $x = \pi/2$, $\phi = \pi/2$. This curve (dashed line in Fig. 7) determines the orientation of a bounce and hence the fundamental domain. Each symbol string $s_{t-1}.s_t$ is here drawn in two disconnected rectangles, but any two rectangles with the same labeling are symmetric to each other.

In the phase space III, Fig. 8, there are three arrows $a.$, $d.$, and $e.$ going into the node in Fig. 5, and there are four outgoing arrows $.a$, $.j$, $.c$, and $.k$. If $x/2 < \phi < \pi/2 + x/2$ the previous bounce was not off the same semi-circle, so such points belong to the phase space II. We then only have to look at $\phi < x/2$. The partition curve separating $a.$ and $d.$ is the line $\phi = x/4$. The curve separating $d.$ and $e.$ is one fold of the unstable manifold of the horizontal 2-cycle (dashed curve). This particular fold consist of all orbits which bounce back and forth between the two semi-circles and then twice off the same semi-circle wall, with the second bounce off the semi-circle yielding a point in the fold. The curve separating $.a$ and $.j$ is the line $\phi = \pi/2 - x/2$, and the curves separating $.j$, $.c$ and $.k$ correspond to the orbits bouncing straight into a singular point of the other semi-circle.

The rectangles map into each other with the partition lines mapping into the partition lines, and in the limit $a \to \infty$ the borders are invariant manifolds, as required for a complete Markov partition.

As the length $a$ decreases, the rectangles change, and at $a = 1$ the rectangle $j.f$ in phase space I and the rectangle $a.k$ in phase space III disappear completely. This is the bifurcation point for the $\overline{011022} = \overline{cea}$ cycle and the associated family of orbits (see ref.[8]) which includes the orbits $\overline{234010} = \overline{jfgdce}$ and $\overline{113020} = \overline{akgdce}$ with points in these rectangles. Also other rectangles partly disappear and become smaller. This is not easily seen in the phase space figures because there the metric area might sometimes grow even while the number of orbits of given length within the rectangle decreases. The point is that the phase space is not a convenient space to use when investigating admissibility of symbolic dynamics orbits. It is much easier to work in a symbol plane, where each orbit occupies a fixed position independent of the parameter $a$.

## 5 Well-ordered symbols and the symbol plane

We now construct a topologically faithful symbol plane representation of each of the three phase spaces introduced above. In the symbol plane any point belonging to any orbit existing for $a = \infty$ maps into a square $(\delta, \gamma)$, with $0 \leq \delta \leq 1$ and $0 \leq \gamma \leq 1$. From now on we will define a point on an orbit by its position in the symbol plane, without bothering to compute its position in the phase space coordinates. The phase space partitions of the preceding section will be needed only to motivate the topologically correct ordering of different symbols in the $(\gamma, \delta)$ plane.

In Fig. 9 (a), (b) and (c) the first generation of the partition is drawn in the $(\delta, \gamma)$ planes for the three phase spaces. Note that the ordering of the different rectangles is the same as the ordering of the corresponding phase spaces rectangles, Figs. 6, 7 and 8. In the symbol plane we simply divide the $\delta$ and $\gamma$ axes into equal intervals for each in and out arrow from the nodes of Fig. 5. In Fig. 9 (a) the vertical $\gamma$-axis is divided in two for the two possible future symbols $.g$ and $.f$: $0 \leq \gamma < 1/2$ corresponds to $.g$, and $1/2 \leq \gamma \leq 1$ corresponds to $.f$. The horizontal $\delta$-axis is divided into five equal intervals for the five possible past symbols $j.$, $h.$, $i.$, $k.$, and $f.$, with the ordering of the five symbols the same as in the phase space I, Fig. 6. The intervals are $0 \leq \delta < 1/5 \rightarrow j.$, $1/5 \leq \delta < 2/5 \rightarrow h.$, $2/5 \leq \delta < 3/5 \rightarrow i.$, $3/5 \leq \delta < 4/5 \rightarrow k.$, and $4/5 \leq \delta \leq 1 \rightarrow f.$.

We then associate an integer $v_t$ to each of the symbols, ordered as the corresponding rectangles in the $(\delta, \gamma)$ plane: $.g \rightarrow 0$ and $.f \rightarrow 1$ for the future and: $j. \rightarrow 0$, $h. \rightarrow 1$, $h. \rightarrow 2$, $k. \rightarrow 3$ and $f. \rightarrow 4$ for the past. This is the first step in constructing *well ordered* symbols; we denote the first future symbol by $v_1$, and the first past symbol by $v_0$.

Figs. 9 (b) and (c) show the corresponding construction for the symbol planes II and III. The symbol plane is divided into rectangles, and the rectangles are labelled with the future and past symbols, ordered as the corresponding phase space rectangles. The well ordered symbols $v_t$ obtained from the symbols $s_t$ for all three nodes are given in table 2. Note that a symbol $s_t$ may correspond to a different symbol $v_t$, depending on whether it is a future or a past symbol.

The next generation of partition is obtained by using the next partition of $\gamma$ and $\delta$ from the node that one moves to (respectively came from) in Fig. 5. We illustrate this by an example.

Consider the rectangle $j.f$ in symbol plane I, Fig. 9 (a). All points in this rectangle correspond to a future symbol $f$. From the Markov graph Fig. 5 we find that $f$ returns to node I and we therefore use the partition of $\gamma$ of Fig. 9 (a). This yields the second generation partition of $\gamma$ into the two intervals $(0.5, 0.75)$ and $(0.75, 1.0)$. All points in the rectangle $j.f$ also correspond to a past symbol $j$, and Fig. 5 shows that the previous node was node III. The partition of $\delta$ is then made according to Fig. 9 (c) which yields three intervals $(0, 1/15)$, $(1/15, 2/15)$ and $(2/15, 1/5)$. The rectangle $j.f$ now splits into a second generation partition of 6 rectangles, with the labeling of each rectangle given in Fig. 10 (a). Note that the ordering of the past symbols $aj.$, $dj.$ and $ej.$ is reversed compared to Fig. 9 (c). If we analyze the second generation partition of the rectangle $k.f$ which also comes from node III, we get three intervals along the $\delta$-axis, but in this case the ordering of Fig. 9 (c) is preserved; $ek.$, $dk.$ and $ak.$. The first and second generations of the partition of the whole symbol plane I are drawn in Fig. 10 (b).

Whether the arrow connecting the nodes in Fig. 5 implies order reversal or preservation has to be incoded in the construction of the symbol plane $(\delta, \gamma)$. It is easy to verify that for future symbols the symbols $c$, $d$, $h$ and $k$ reverse the ordering, while $e$, $h$ and $j$ reverse the ordering for the past symbols. The remaining symbols preserve the ordering. The ordering can be reversed because bounces off the straight wall reverse the ordering of two neighboring orbits, while the bounces off the semi-circles preserve the ordering.

We can now calculate the symbol plane coordinates $(\gamma, \delta)$ of any orbit given its symbol string representation. The algorithm for computing the values $\gamma$ and $\delta$ is more complicated than for a simple horseshoe map[4] because here the symbol plane is partitioned into rectangles of different sizes (see Fig. 10 (b)).

Let $A_i^{\text{in}}$ and $A_i^{\text{out}}$ be the number of arrows into and out from the node $i$: $A_I^{\text{in}} = 5$, $A_I^{\text{out}} = 2$, $A_{II}^{\text{in}} = 3$, $A_{II}^{\text{out}} = 5$, $A_{III}^{\text{in}} = 3$ and $A_{III}^{\text{out}} = 4$. The orbit is given by the symbol string $\cdots s_{-1}s_0.s_1s_2\cdots$. Let $i_t$ be the node to which the arrow $s_t$ arrives at for $t \leq 0$, or from which it leaves for $t > 0$. Let $V^- = \cdots v_{-2}v_{-1}v_0$ be given by the past values $v$ from Table 2 obtained from the symbols $\cdots s_{-2}s_{-1}s_0$, and $V^+ = v_1v_2v_3\cdots$ be given by the future values $v$ obtained from the symbols $s_1s_2s_3\cdots$.

The "past" coordinate $\delta$ is given by $\delta = \lim_{t\to-\infty} \delta_t$, where $\delta_t$ with $t \le 1$ is computed iteratively by

$$
\begin{aligned}
m_{t-1} &= m_t/A^{\text{in}}_{i_{t-1}}\,, && m_1 = 1.0 \\
p_{t-1} &= \begin{cases} p_t & \text{if} \quad s_t \in \{a,b,c,d,f,g,i,k\} \\ -p_t & \text{if} \quad s_t \in \{e,h,j\} \end{cases}\,, && p_0 = 1 \\
\delta_{t-1} &= \delta_t + \begin{cases} v_{t-1}m_{t-1} & \text{if} \quad p_{t-1} = 1 \\ (A^{\text{in}}_{i_{t-1}} - 1 - v_{t-1})m_{t-1} & \text{if} \quad p_{t-1} = -1 \end{cases}\,, && \delta_1 = 0.0\,.
\end{aligned}
\tag{3}
$$

The "future" coordinate $\gamma$ is given by $\gamma = \lim_{t\to\infty} \gamma_t$ where $\gamma_t$ with $t \ge 0$ is computed iteratively by

$$
\begin{aligned}
m_{t+1} &= m_t/A^{\text{out}}_{i_{t+1}}\,, && m_0 = 1.0 \\
p_{t+1} &= \begin{cases} p_t & \text{if} \quad s_t \in \{a,b,e,f,g,i,j\} \\ -p_t & \text{if} \quad s_t \in \{c,d,h,k\} \end{cases}\,, && p_1 = 1 \\
\gamma_{t+1} &= \gamma_t + \begin{cases} v_{t+1}m_{t+1} & \text{if} \quad p_{t+1} = 1 \\ (A^{\text{out}}_{i_{t+1}} - 1 - v_{t+1})m_{t+1} & \text{if} \quad p_{t+1} = -1 \end{cases}\,, && \delta_0 = 0.0\,.
\end{aligned}
\tag{4}
$$

# 6 The pruning front

We are now finally in position to draw the pruning front and determine the inadmissible orbits for a finite length stadium. This way of describing symbolic dynamics of 2-*d* hyperbolic systems was introduced by Cvitanović *et al.* for the Hénon map[4, 5], and applied to dispersive billiards in ref. [8]. The singular point on the border determines whether a given symbol sequence corresponds to an admissible orbit. All orbits which are pruned as the stadium length decreases disappear as a bounce in the orbit hits the singular point.

The pruning front is constructed by scanning through all orbits starting (or ending) at the singular point at different angles, and mapping the corresponding future and past symbol sequences into the symbol planes I, II and III by algorithms (3) and (4). The resulting pruning fronts, Figs. 11, 12 and 13, are fractal sets of points in the $(\delta, \gamma)$ plane: the area outside the pruning front, the primary forbidden region, contains points corresponding to all inadmissible orbits. In Fig. 14 points belonging to several long chaotic orbits are plotted in the symbol plane I; as expected, the pruning front in Fig. 11 is the border between these points and the primary forbidden region. All white regions in this figure correspond to the forbidden symbol sequences, but one needs to define only the primary forbidden regions as all other regions are images or preimages of these.

The pruning front is monotone in the symbol plane since the symbol plane is constructed with well ordered symbols which respect the foliation of the stable/unstable manifolds.

The primary forbidden region is rather small for $a = 5$. It increases with decreasing $a$ and is already quite large for $a = 0.5$, see Fig. 14. In the integrable limit $a \to 0$ only the rotation orbits existing in the circle survive[16]. For small $a$ the pruning front description is still correct but probably not convenient for calculations.

We now approximate the pruning front by partitioning the symbol plane into an integer lattice, and tracing out approximate pruning fronts along the lattice lines of this partition. In Fig. 15 the symbol plane I partition lines of the first and second generation are drawn together with the pruning front. The rectangles completely in the primary forbidden regions are shaded and correspond to the symbol substrings $j.f$, $f.gd$, $aj.ge$, $gh.ff$, $ch.ff$, $kf.gh$, $ff.gh$ and $ak.gd$. From the other symbol planes we get further forbidden substrings: all completely forbidden substrings up to length 4 are

$$\begin{array}{l} jf,\ ak \\ fgd,\ gea \\ ajge,\ ghff,\ chff,\ kfgh,\ ffgh,\ cdkf,\ gdkf,\ aace,\ acea,\ kgej,\ fgej. \end{array} \qquad (5)$$

Given such list we can construct a Markov diagram which generates all admissible orbits in this approximation. As only the fully pruned rectangles have been removed, this approximation underestimates the number of pruned orbits.

# 7 An application: construction of topological zeta functions

The Markov diagrams can be applied to calculating averages and spectra of classical and quantum mechanical systems. As an illustration, we now determine the topological entropy $h$ in a few simple approximations. The topological entropy is a measure of how fast the number of periodic orbits grows as the cycle length increases. The number of periodic orbits with symbol string length $l$ in the limit $l \to \infty$ is given by $N(l) \sim e^{hl}$, where the value of $h$ is given by the logarithm of the inverse of the smallest zero of the topological zeta function[21].

For finite Markov graph the topological zeta function is given by the characteristic polynomial of the graph. The characteristic polynomial is obtained by counting the number of non-self intersecting closed paths on the graph, and combinations of such closed paths without common nodes[5].

The characteristic polynomial for the $a = \infty$ graph of Fig. 5 is

$$1/\zeta_{\mathrm{top}}(z) = 1 - 3z - z^2 - z^3 \,, \tag{6}$$

with leading eigenvalue $z = e^h$,

$$h = \ln 3.38298\ldots = 1.21875\ldots$$

This is the upper bound on the topological entropy of any stadium billiard.

The Markov diagram for two forbidden strings of length 2 approximation to the $a = 1$ symbolic dynamics, Fig. 16 yields

$$1/\zeta_{\mathrm{top}}(z) = 1 - 3z - z^2 - z^3 + 4z^4$$

with the slightly smaller topological entropy

$$h = \ln 3.28428\ldots = 1.18915\ldots$$

The Markov diagram with the four forbidden strings of lengths 2 and 3 from the list (5) yields

$$1/\zeta_{\mathrm{top}}(z) = 1 - 3z - 3z^3 + 5z^4 + 2z^5 + 6z^6 + z^7$$

with the even smaller topological entropy

$$h = \ln 3.10061\ldots = 1.131600\ldots$$

For comparison, the numbers of points in cycles listed in the table I of ref. [15] for the $a = 1.6$ yields $h \approx 1.1$.

As more and more forbidden sequences are taken into account, the construction of topological zeta functions becomes more laborious, but not impossible. The most detailed evaluation of topological entropy from a pruning front has been implemented by Grassberger *et al.*[22] for the Hénon attractor. However, while in the case of almost purely hyperbolic systems such as the Hénon attractors and repellers, organization of cycles by finite alphabet symbolic dynamics also reflects their relative importance in evaluation of chaotic averages, for nonhyperbolic systems finite alphabet symbolic dynamics is rather less useful, as it does not account correctly for itermittency effects.

# 8 Conclusions

We have introduced a symmetry reduced symbolics dynamics for the stadium billiard, obtained an exact description of all admissible orbits in terms of a pruning front, and shown how to construct approximate finite Markov partitions of the stadium phase space. The symbolic dynamics is a slight improvement of the Biham-Kvale description, and the construction of the Markov graph for an infinitely long stadium and a pruning front for a finite length stadium are new results. While the stadium billiard is one of the most commonly used examples of an ergodic dynamical system, its symbolic dynamics is more complicated than that of other systems analyzed in detail in literature, such as the $n$-disk pinballs and the Hénon map. Nevertheless, it is possible to obtain useful finite approximations to the symbolic dynamics, and guarantee that all periodic orbits up to given length have been taken into account. We have concentrated here on purely topological description of the dynamics, and have not attempted any measure dependent periodic orbit calculations. Depending on the quantity computed, those might suffer from the usual ills of nonhyperbolic dynamical systems, such as intermittency effects due to the presence of bouncing ball and whispering gallery orbits.

KTH gratefully acknowledge financial support by the A. von Humboldt foundation and the Norwegian Research Council.

| $s_t$ | $\sigma_{t-1}\sigma_t$ | $\sigma_{t-n-1}0^n\sigma_t$ |
|---|---|---|
| $a$ | 11 | |
| | 22 | |
| $b$ | 00 | |
| $c$ | 10 | |
| | 20 | |
| $d$ | 01 | $_10^n1_$ |
| | 01 | $_40^n1_$ |
| | 02 | $_20^n2_$ |
| | 02 | $_30^n2_$ |
| $e$ | 01 | $_20^n1_$ |
| | 01 | $_30^n1_$ |
| | 02 | $_10^n2_$ |
| | 02 | $_40^n2_$ |
| $f$ | 34 | |
| | 43 | |

| $s_t$ | $\sigma_{t-1}\sigma_t$ | $\sigma_{t-n-1}0^n\sigma_t$ |
|---|---|---|
| $g$ | 30 | |
| | 40 | |
| $h$ | 03 | $_20^n3_$ |
| | 03 | $_30^n3_$ |
| | 04 | $_10^n4_$ |
| | 04 | $_40^n4_$ |
| $i$ | 03 | $_10^n3_$ |
| | 03 | $_40^n3_$ |
| | 04 | $_20^n4_$ |
| | 04 | $_30^n4_$ |
| $j$ | 23 | |
| | 14 | |
| $k$ | 24 | |
| | 13 | |

Table 1: Definition of the reduced symmetry 1/4 stadium symbols $s$ from the symbols $\sigma$. The second column defines the orientation dependent symbols by indicating the last symbol that preceeds a string of 0's.

| $s$ | $v$ |
|---|---|
| $.g$ | 0 |
| $.f$ | 1 |
| $j.$ | 0 $(r)$ |
| $h.$ | 1 $(r)$ |
| $i.$ | 2 |
| $k.$ | 3 |
| $f.$ | 4 |

| $s$ | $v$ |
|---|---|
| $.d$ | 0 $(r)$ |
| $.h$ | 1 $(r)$ |
| $.b$ | 2 |
| $.i$ | 3 |
| $.e$ | 4 |
| $b.$ | 0 |
| $c.$ | 1 |
| $g.$ | 2 |

| $s$ | $v$ |
|---|---|
| $.k$ | 0 $(r)$ |
| $.c$ | 1 $(r)$ |
| $.j$ | 2 |
| $.a$ | 3 |
| $e.$ | 0 $(r)$ |
| $d.$ | 1 |
| $a.$ | 2 |

Table 2: Definition of the well-ordered symbols $v_t$ from the symbols $s_t$ for the three phase spaces I, II and III. The letter $r$ indicates that this symbol reverses the ordering of the following (or preceding) symbols.

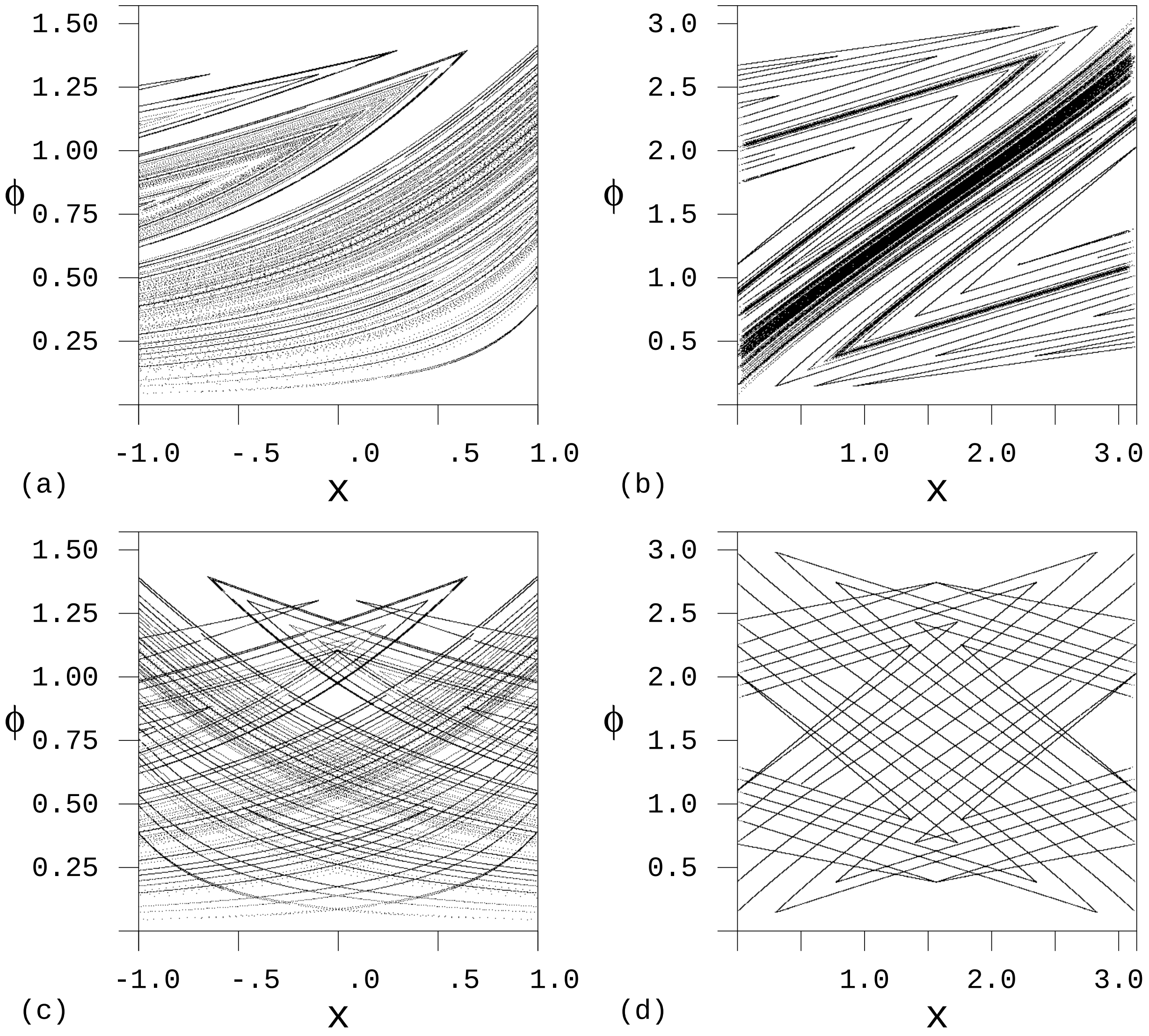

Figure 1: The unstable manifold of the unstable period 2 cycle in the $a = 1.0$ stadium billiard drawn in the $x$ (position) and $\phi$ (outgoing angle) phase space for (a) the straight wall Poincaré section and (b) the semi-circle Poincaré section. Both the stable and the unstable manifolds are drawn in (c) and (d) for the respective phase spaces.

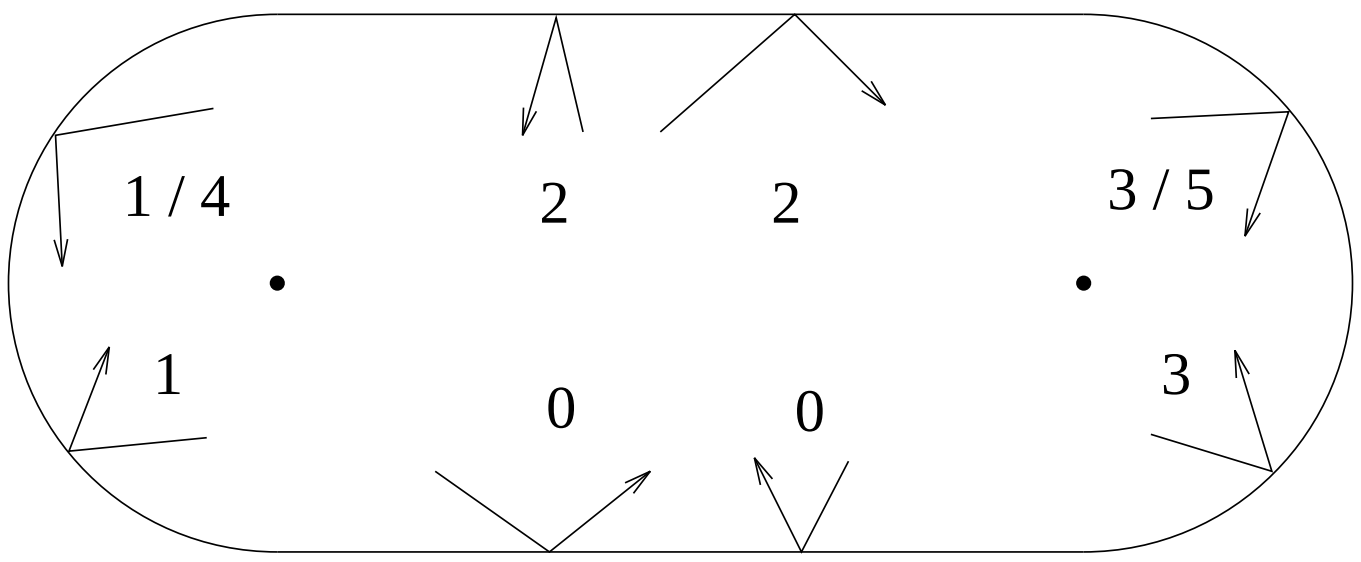

Figure 2: The Biham-Kvale symbols $S_t$.

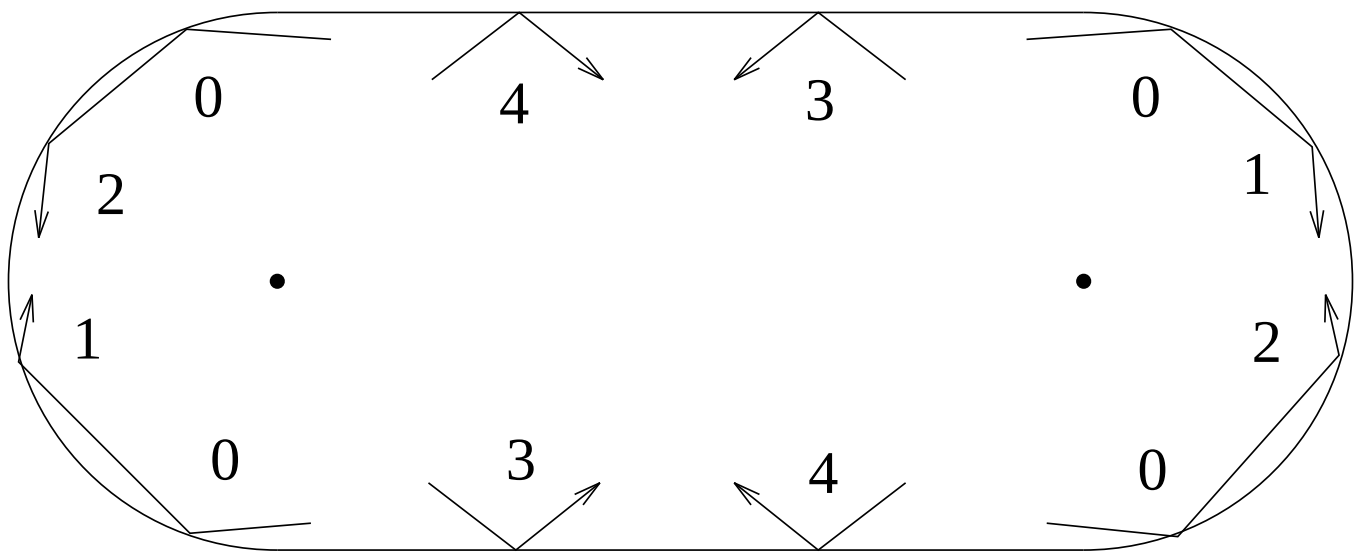

Figure 3: The symbols $\sigma_t$ in the stadium.

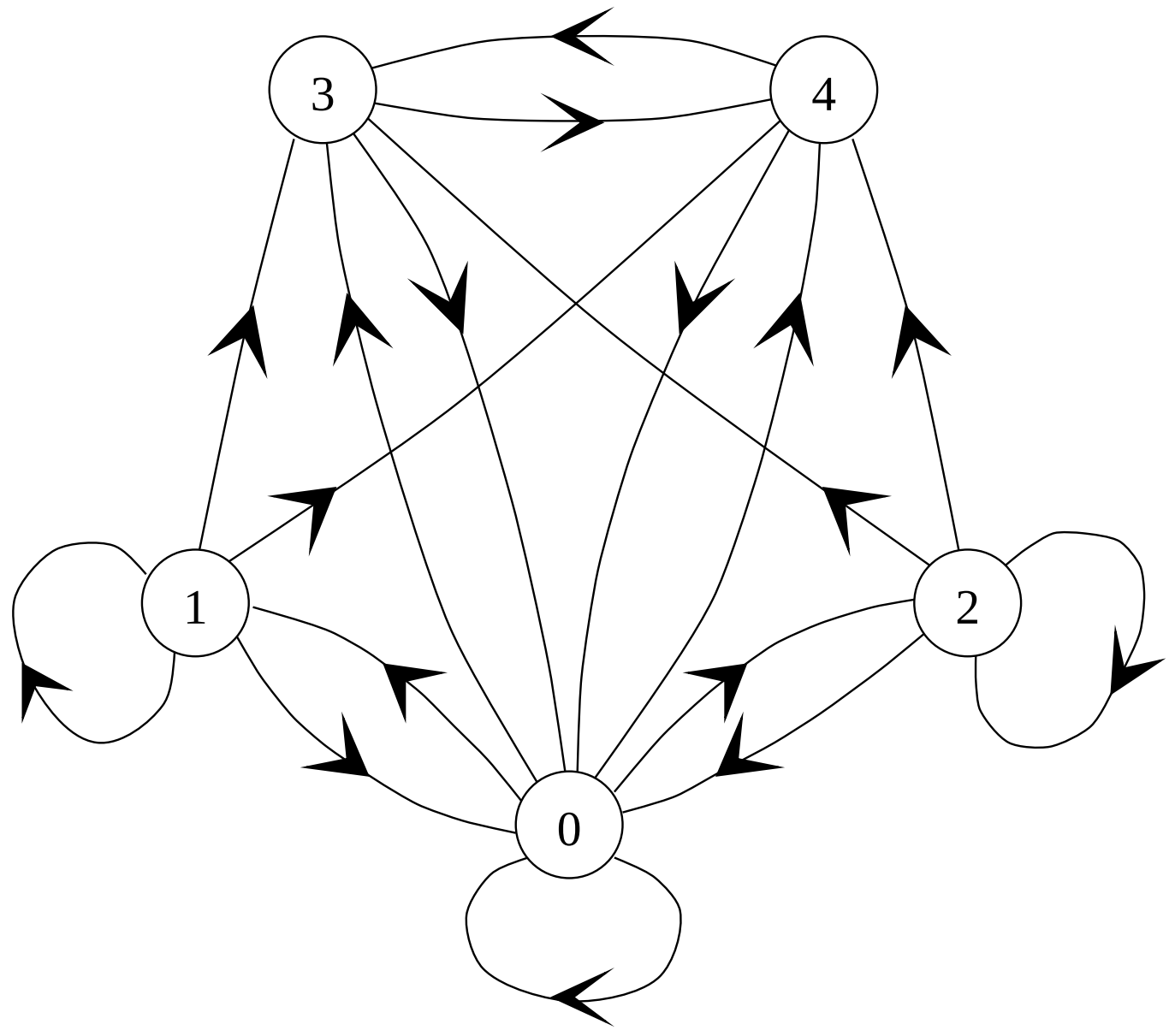

Figure 4: The Markov graph for the $a \to \infty$ infinite length $1/2$ stadium with alphabet $\sigma$.

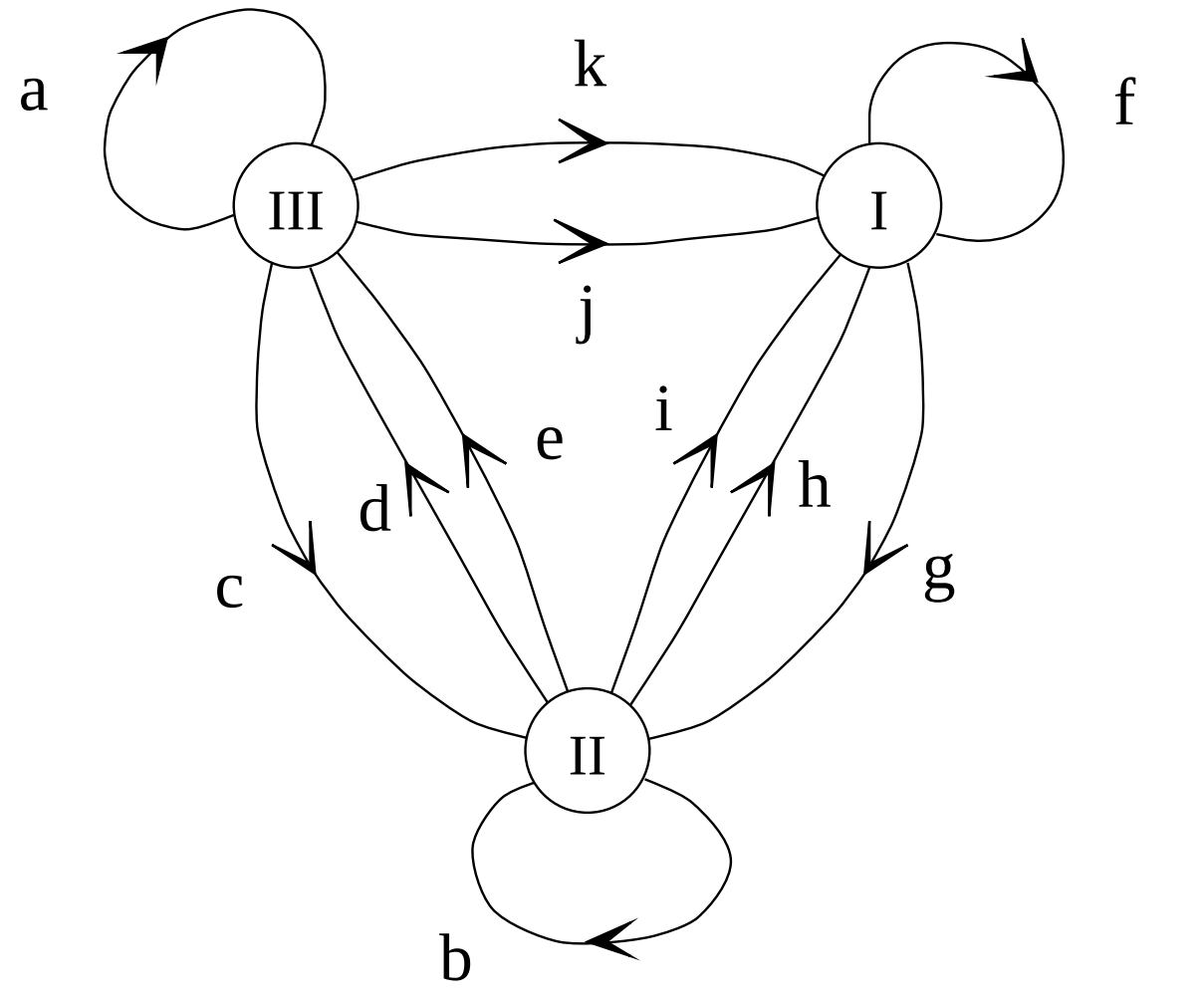

Figure 5: The symmetry reduced Markov graph for the $a \to \infty$ infinite length$1/4$ stadium with alphabet $s$.

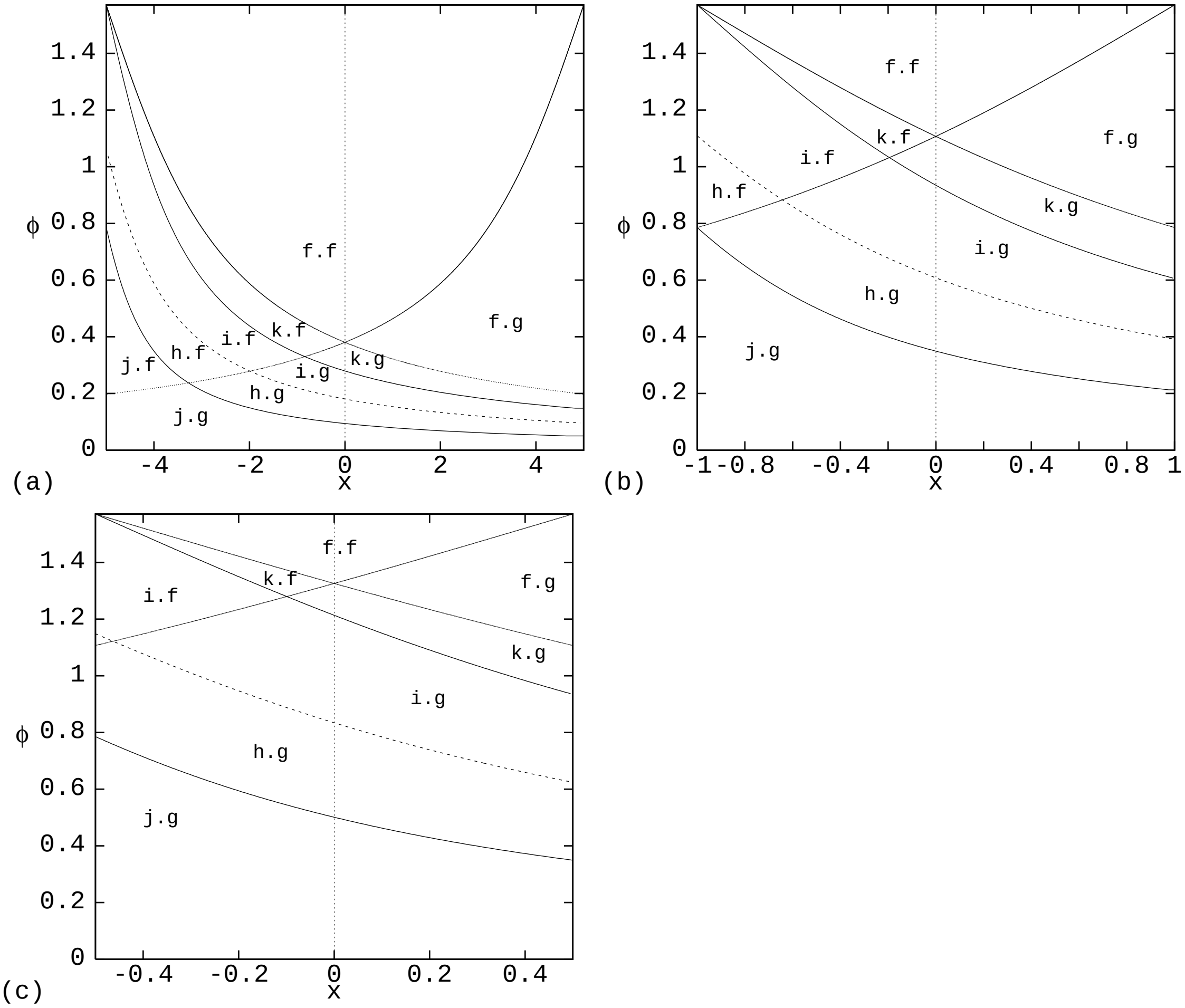

Figure 6: The $a \to \infty$ Markov partition of the phase space I for the straight wall bounces: a) $a = 5$, b) $a = 1$, c) $a = 0.5$.

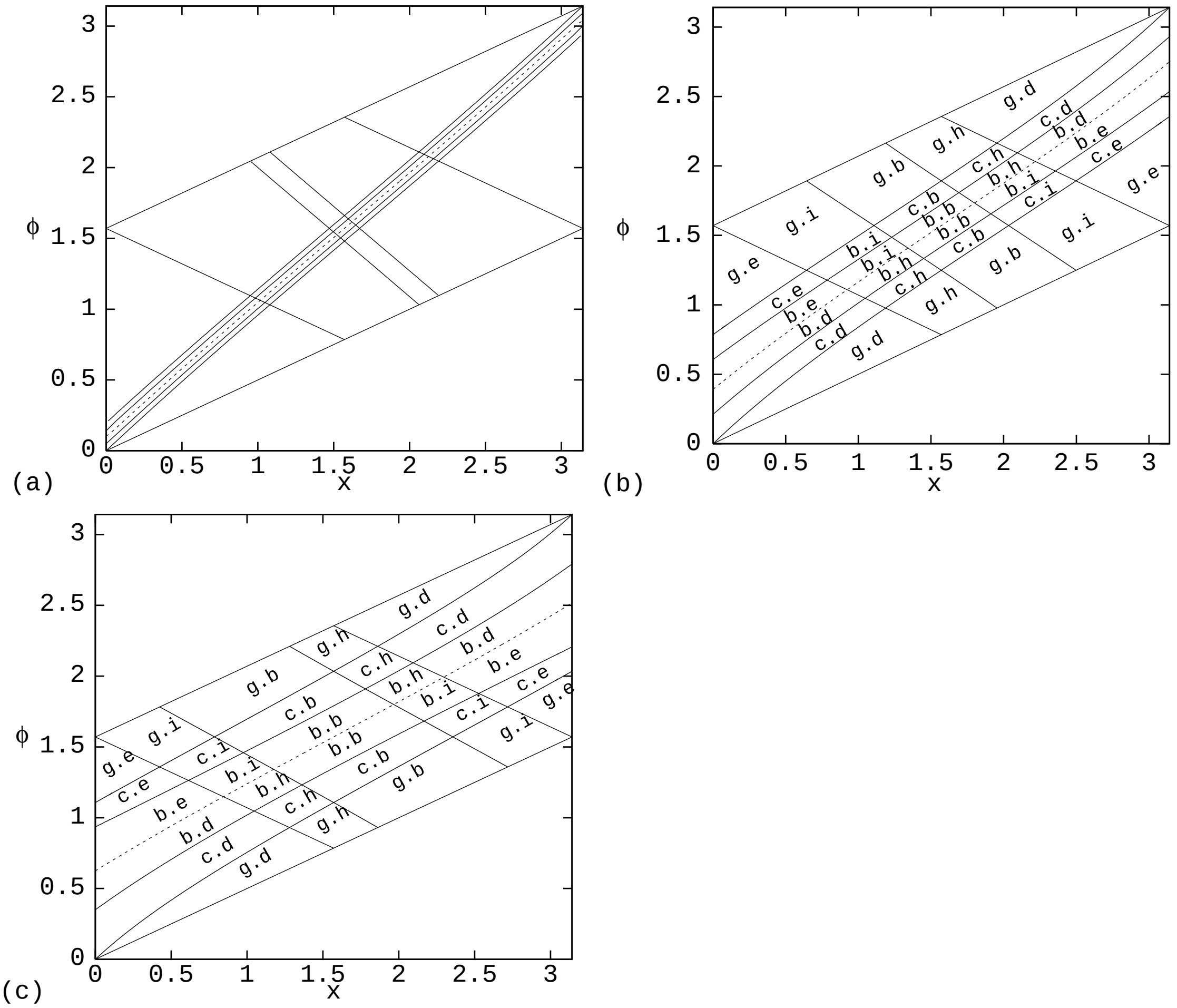

Figure 7: The $a \to \infty$ Markov partition of the first semi-circle bounce phase space II: a) $a = 5$, b) $a = 1$, c) $a = 0.5$.

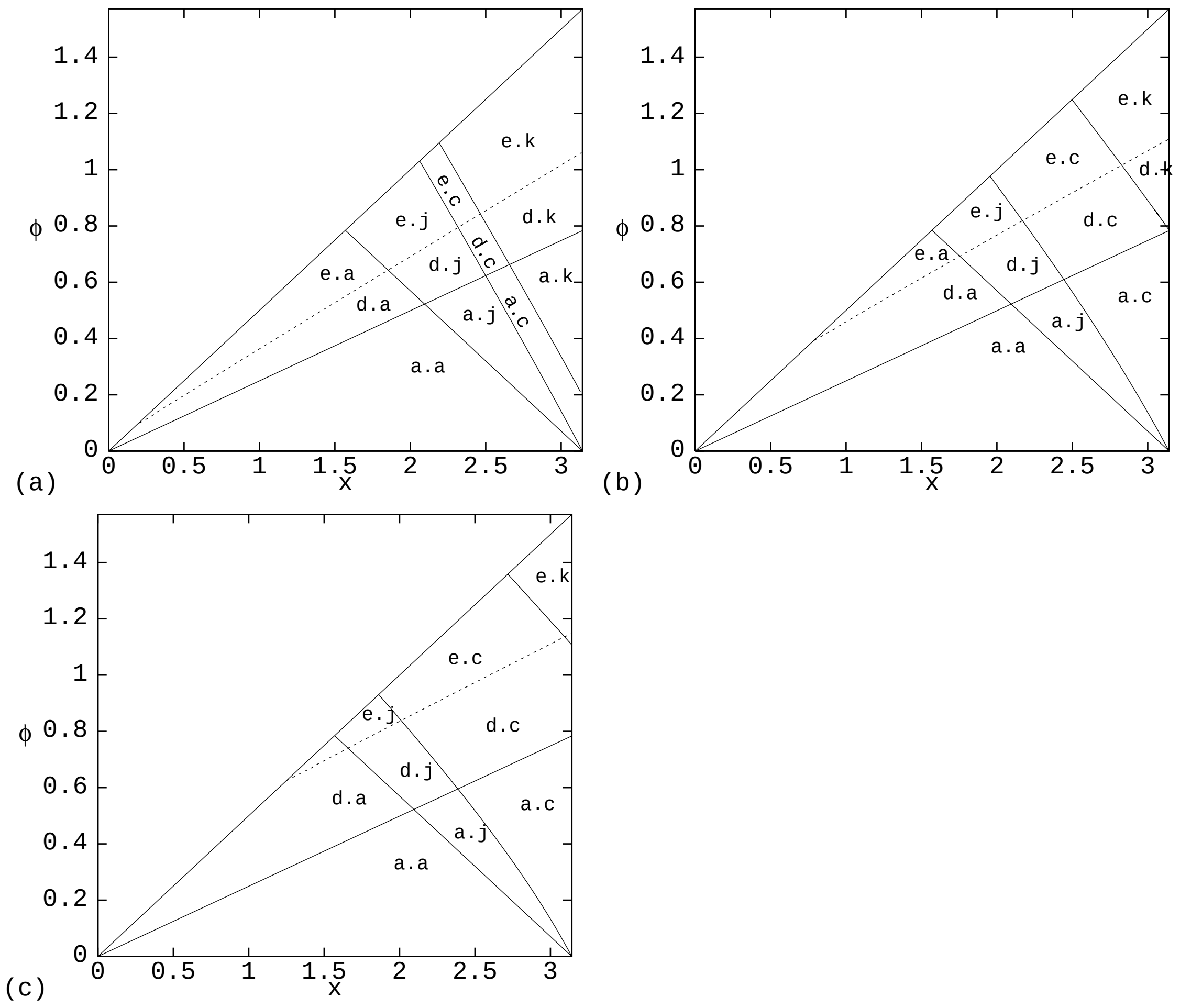

Figure 8: The $a \to \infty$ Markov partition of the "not-first" semi-circle bounce phase space III: a) $a = 5$, b) $a = 1$, c) $a = 0.5$.

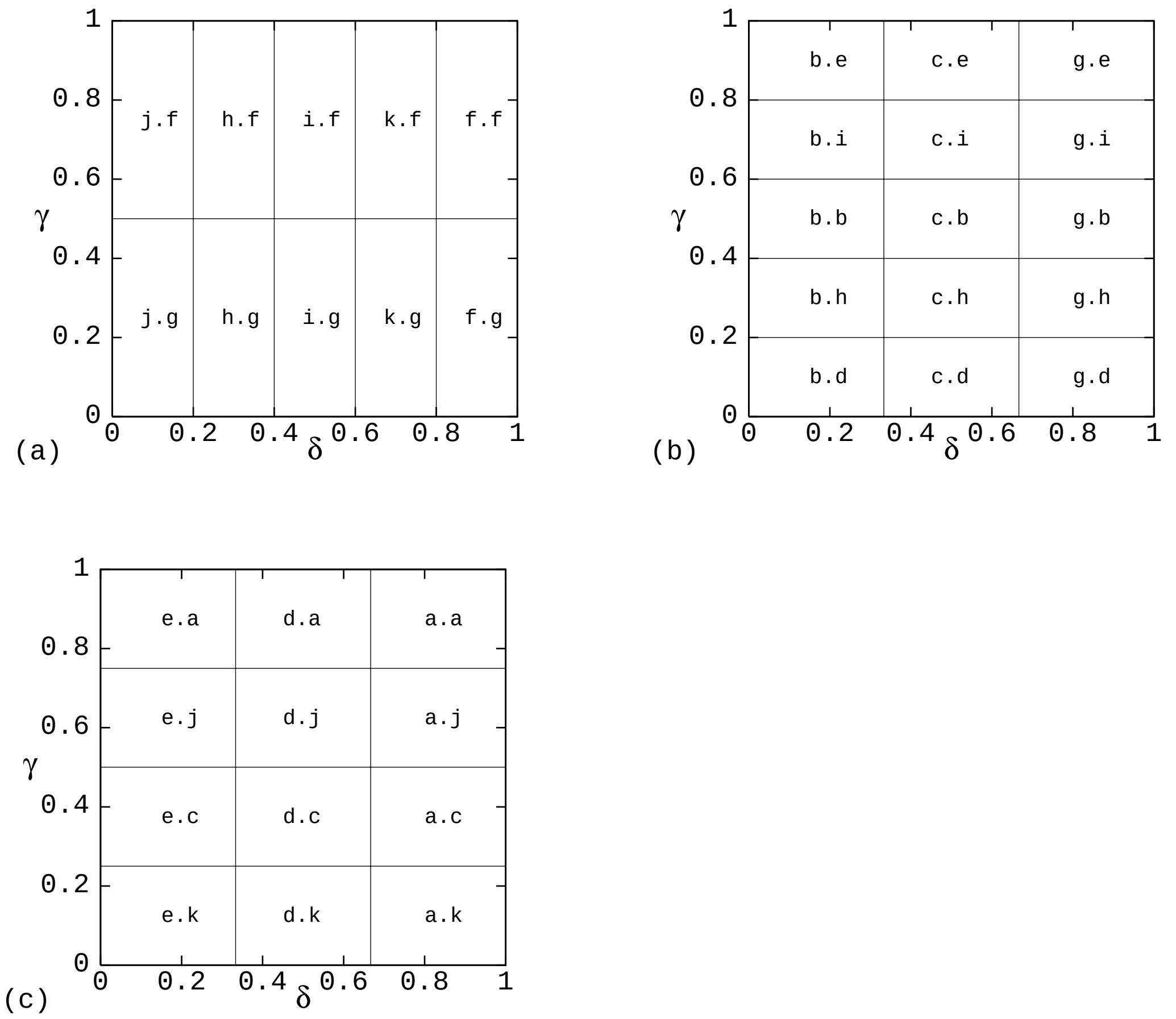

Figure 9: The first level of the partition of the symbol planes (a) I, (b) II, (c) III.

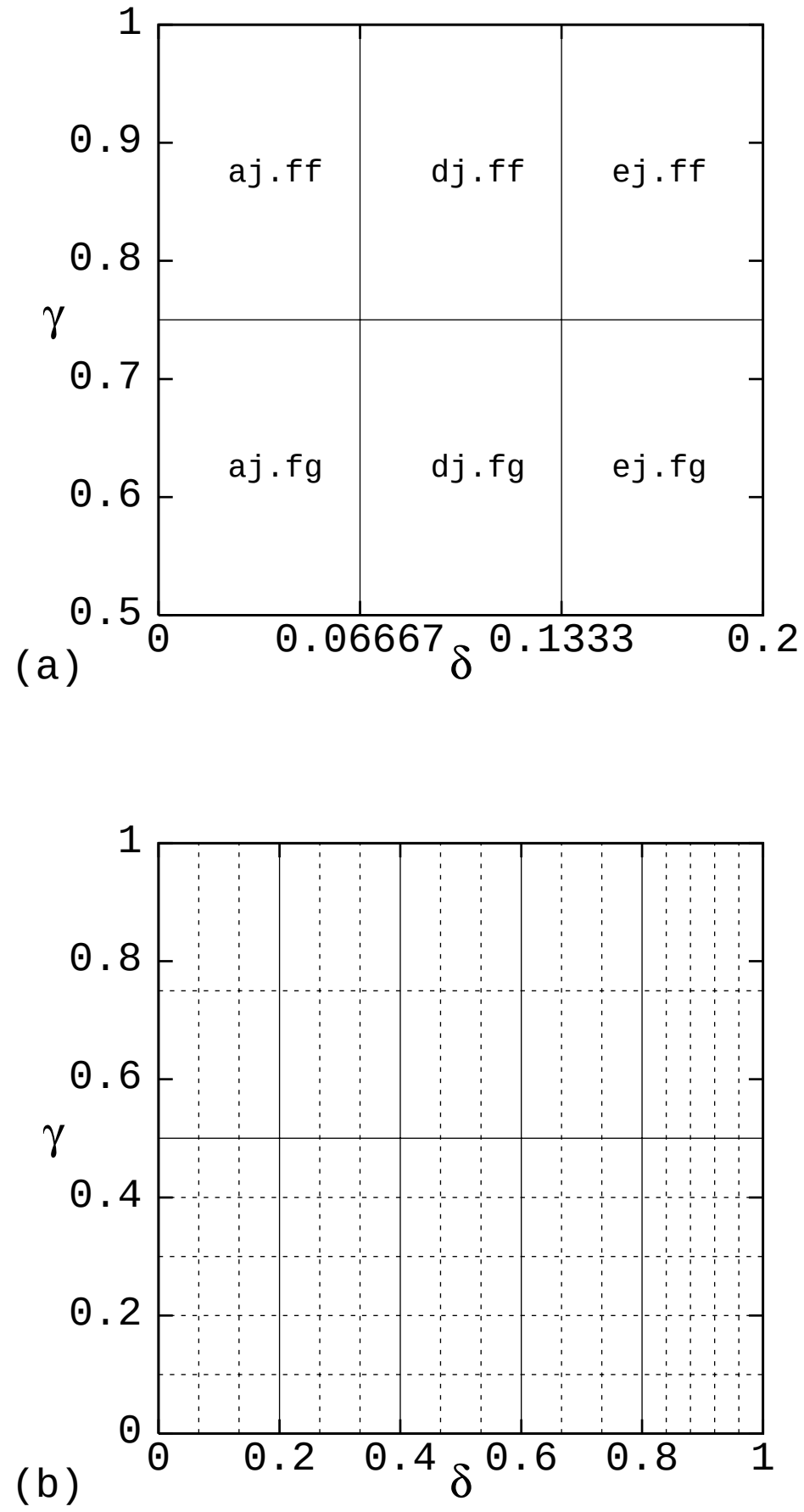

Figure 10: The second level of the partition of the symbol plane I. (a) The rectangle j.f, labeled. (b) The whole plane.

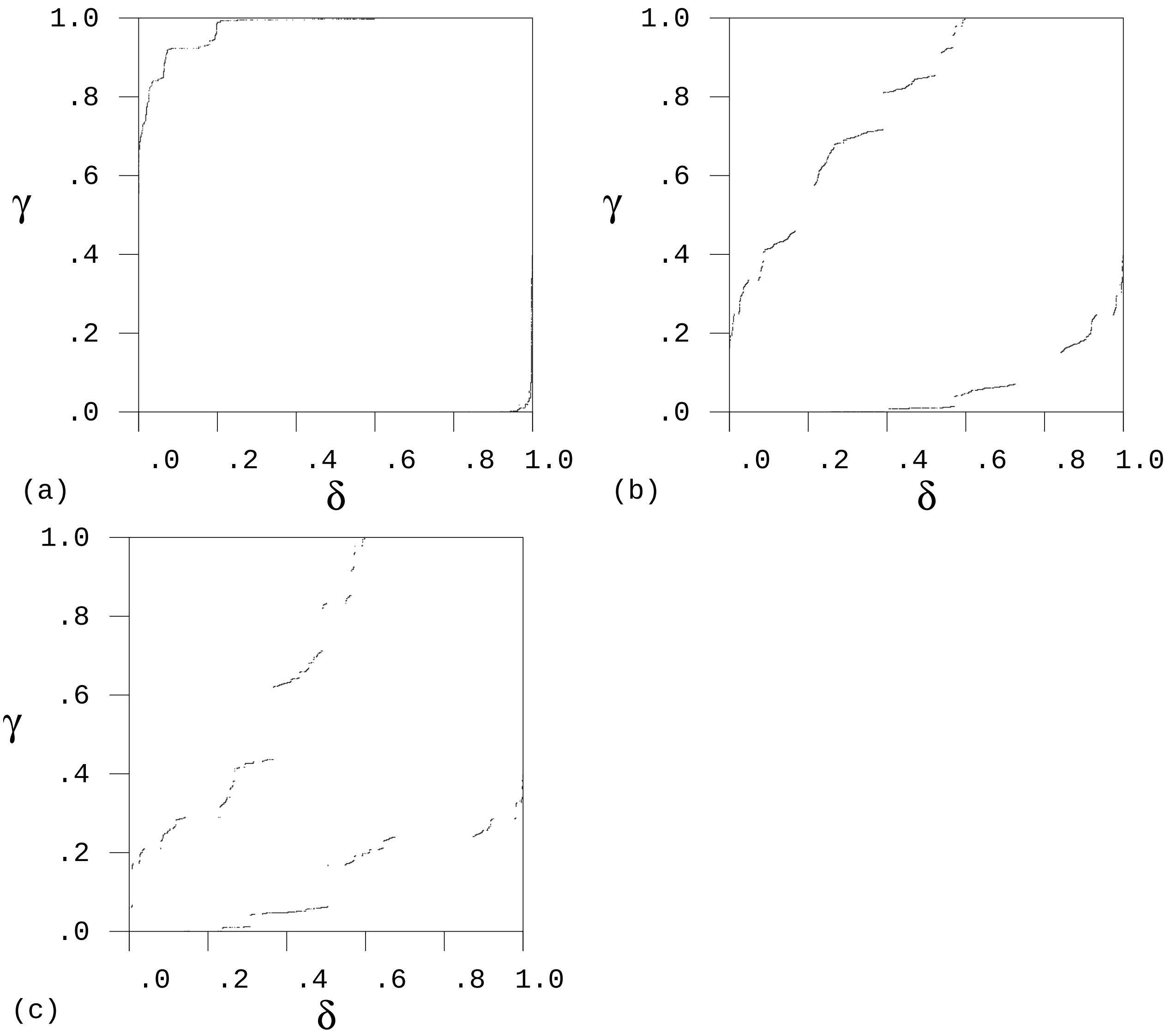

Figure 11: The pruning fronts in the symbol plane I for parameter (a) $a = 5$, (b) $a = 1$ and (c) $a = 0.5$.

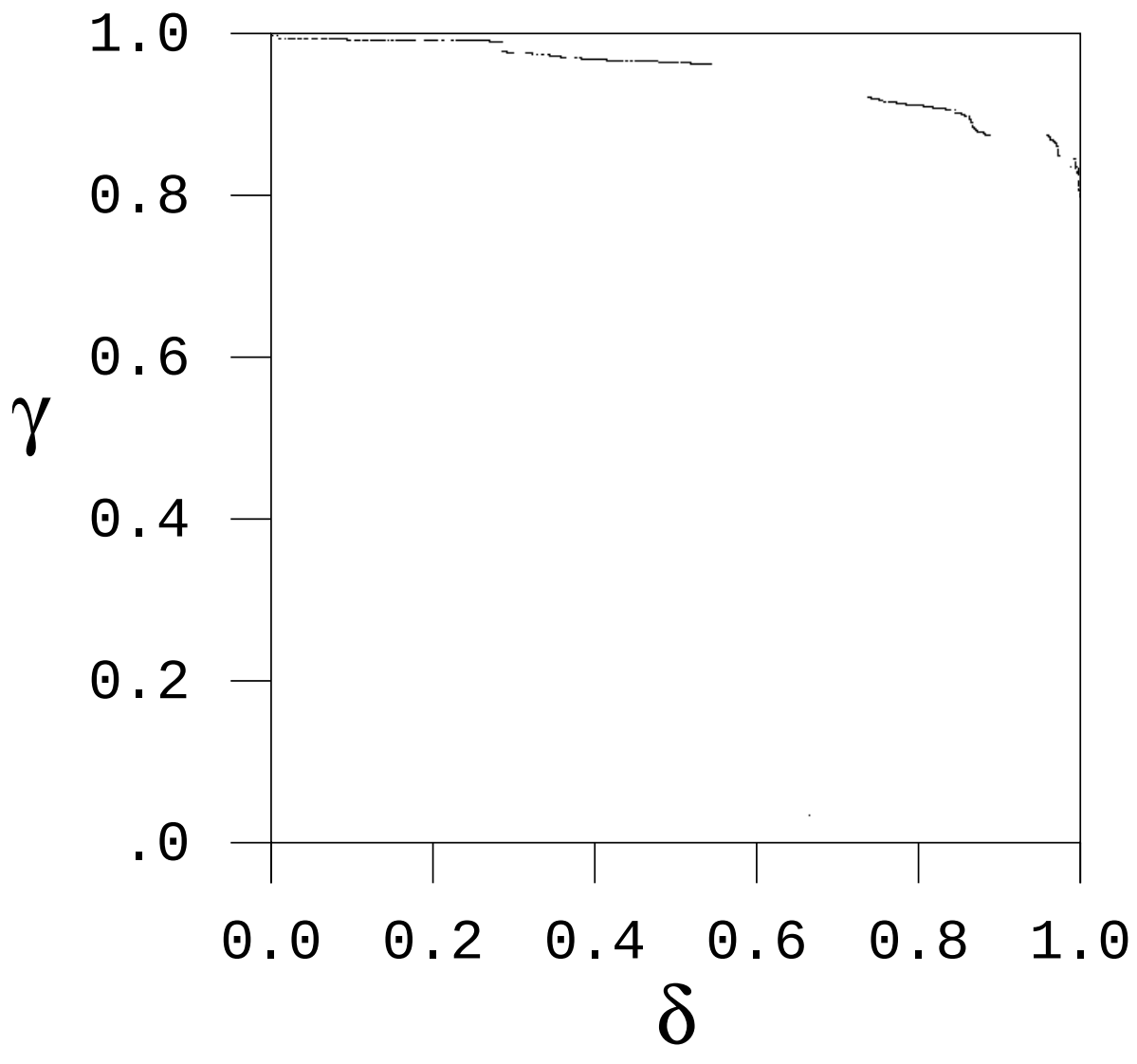

Figure 12: The pruning front in the symbol plane II for parameter $a = 1$.

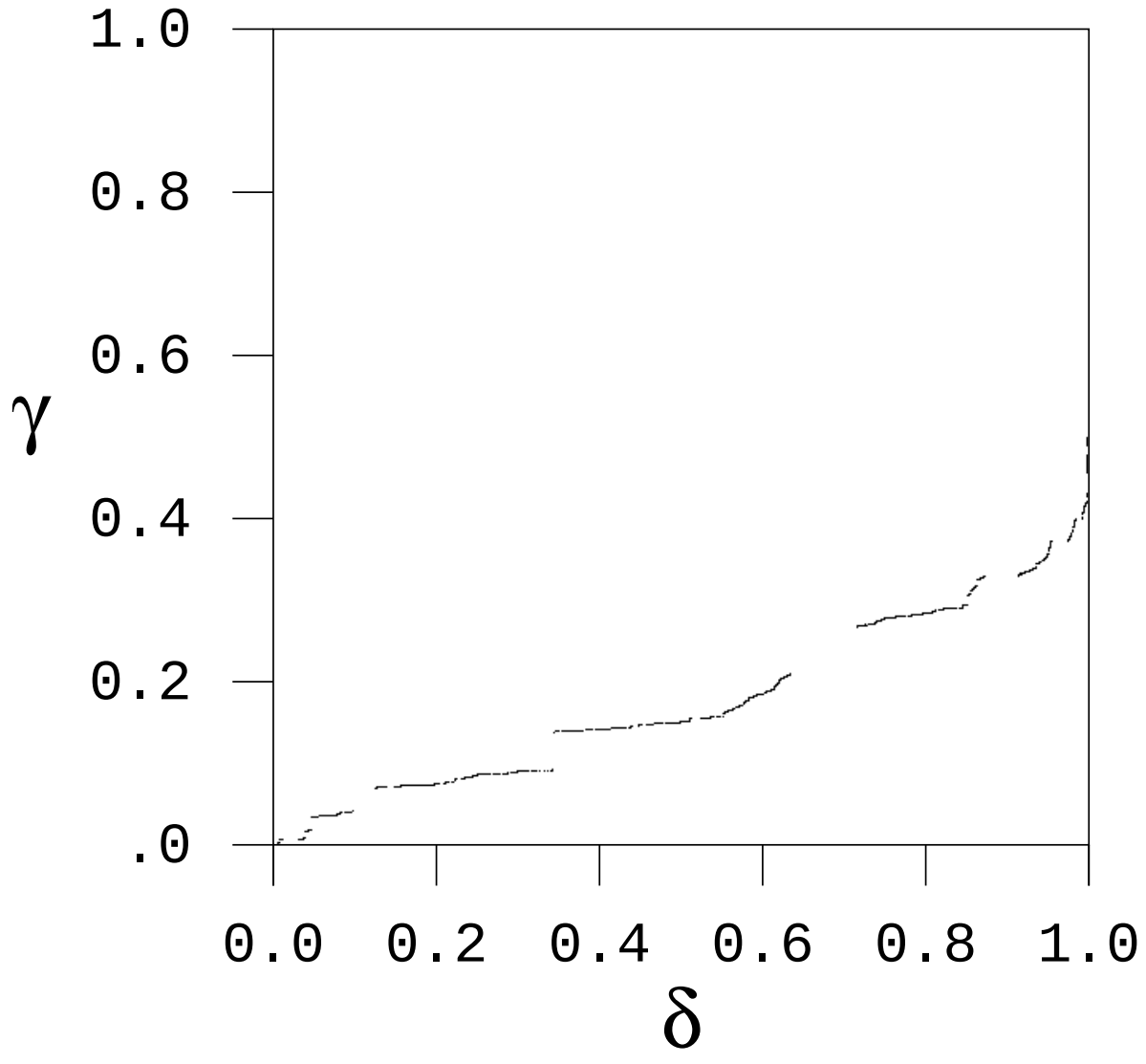

Figure 13: The pruning front in the symbol plane III for parameter $a = 1$.

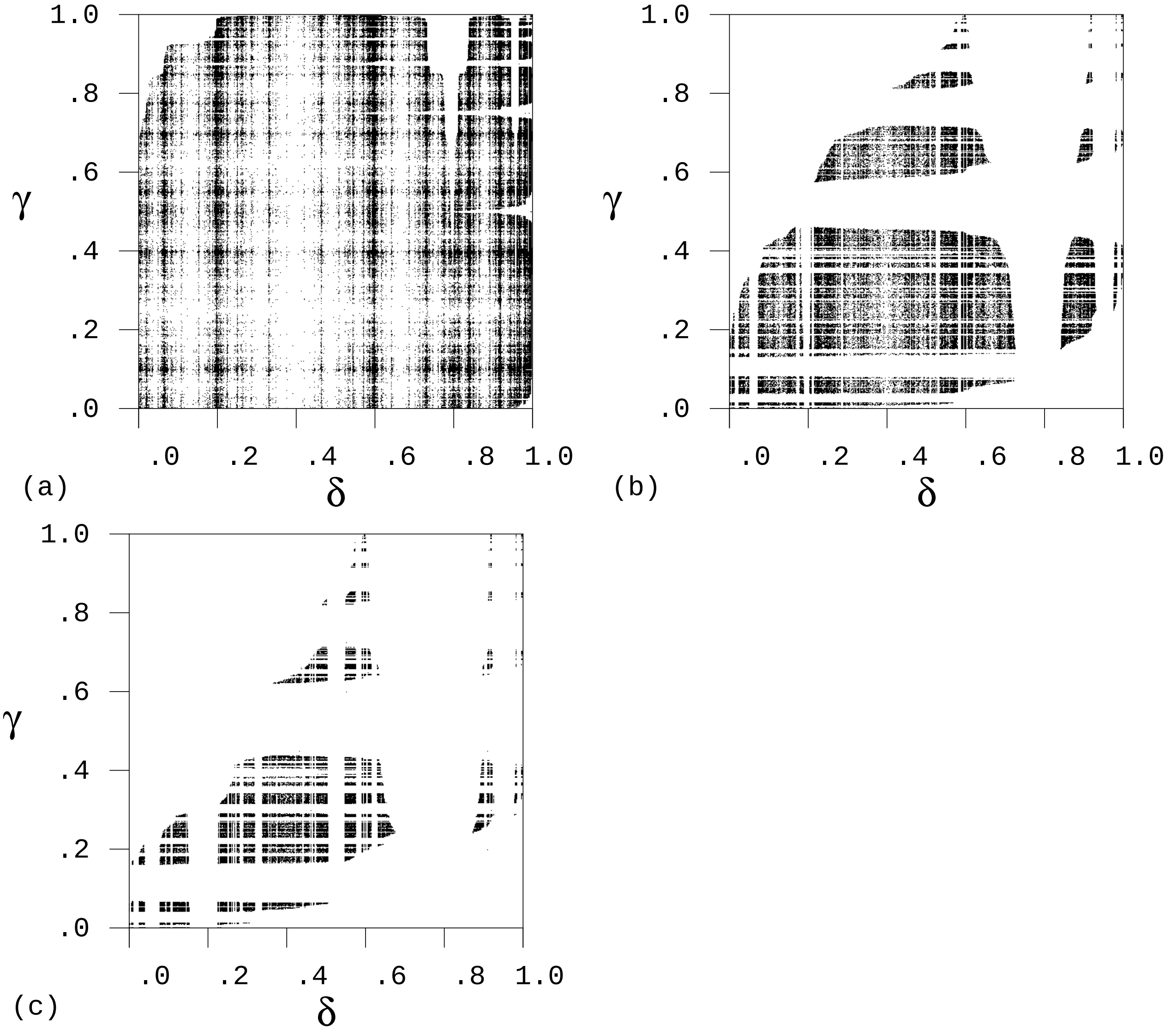

Figure 14: Points in a few randomly initiated chaotic stadium trajectories in the symbol plane I for parameter (a) $a = 5$, (b) $a = 1$ and (c) $a = 0.5$. The primary forbidden regions, delineated by the pruning fronts of Fig. 11, together with their images and preimages, are never visited.

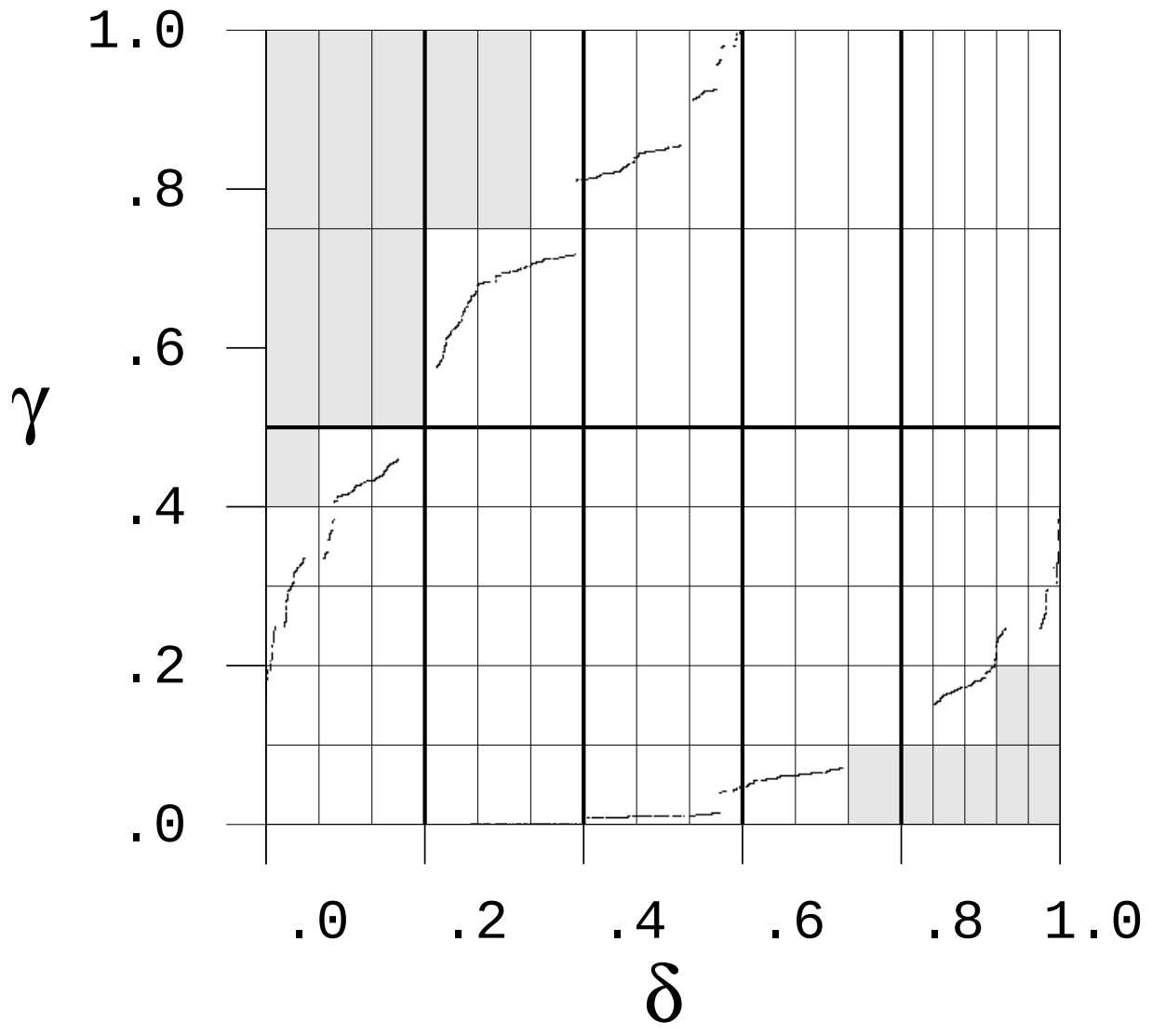

Figure 15: The pruning front in the symbol plane I for parameter $a = 1$, together with the second generation partition lines of Fig. 10 (b). The approximate primary pruned region is shaded.

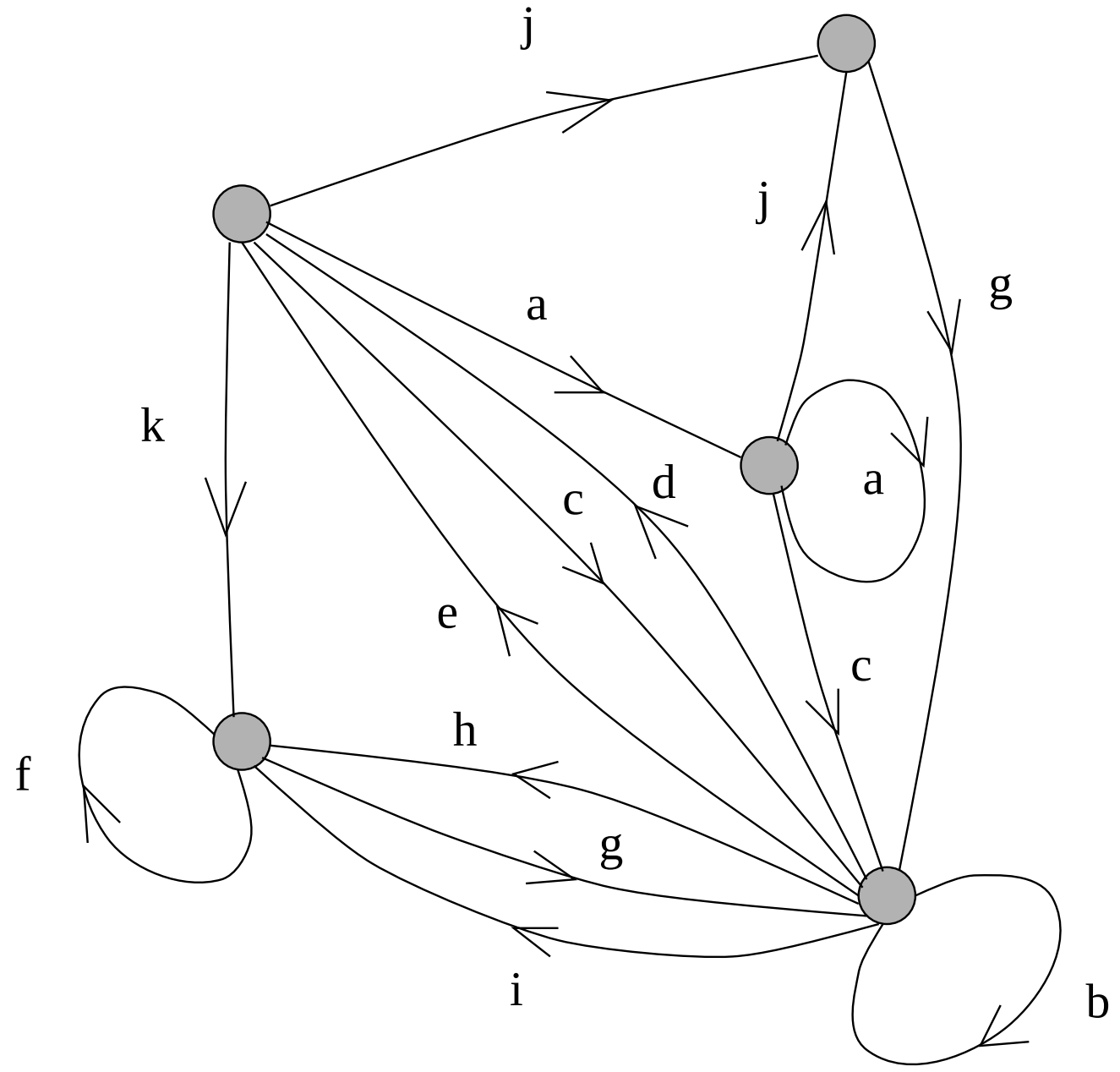

Figure 16: The Markov graph as in Fig. 5, but approximating $a = 1$ finite $1/4$ stadium symbolic dynamics, with the length-2 strings $_jf_$, $_ak_$ of (5) forbidden.